\documentclass[dvips]{article}
\usepackage{graphicx}
\usepackage{amssymb}
\usepackage{graphicx}
\usepackage{dcolumn}
\usepackage{amsmath}
\usepackage{lscape}
\usepackage{longtable}
\usepackage{subfigure}
\usepackage{color}

\begin{document}
\textwidth 16cm
\newcommand{\bd}{\begin{document}}
\newcommand{\ed}{\end{document}}
\newcommand{\bc}{\begin{center}}
\newcommand{\ec}{\end{center}}
\newcommand{\bfr}{\begin{flushright}}
\newcommand{\efr}{\end{flushright}}
\newcommand{\lt}{\left}
\newcommand{\rt}{\right}
\newcommand{\vs}{\vspace}
\newcommand{\hs}{\hspace}
\newcommand{\beq}{\begin{equation}}
\newcommand{\eeq}{\end{equation}}
\newcommand{\lb}{\linebreak}
\newcommand{\pb}{\pagebreak}
\newcommand{\mb}{\makebox}
\newcommand{\fb}{\framebox}
\newcommand{\mc}{\multicolumn}
\newcommand{\ben}{\begin{enumerate}}
\newcommand{\een}{\end{enumerate}}
\newcommand{\bit}{\begin{itemize}}
\newcommand{\eit}{\end{itemize}}
\newcommand{\ol}{\overline}
\newcommand{\un}{\underline}
\newcommand{\lefq}{\lefteqn}
\newcommand{\ba}{\begin{array}}
\newcommand{\ea}{\end{array}}
\newcommand{\beqa}{\begin{eqnarray}}
\newcommand{\eeqa}{\end{eqnarray}}
\newcommand{\beqas}{\begin{eqnarray*}}
\newcommand{\eeqas}{\end{eqnarray*}}
\newcommand{\bfg}{\begin{figure}}
\newcommand{\efg}{\end{figure}}
\newcommand{\bds}{\begin{displaymath}}
\newcommand{\eds}{\end{displaymath}}
\newcommand{\btb}{\begin{tabbing}}
\newcommand{\etb}{\end{tabbing}}
\bc {\huge  Supersymmetric Analysis of the Dirac-Weyl Operator within $\mathcal{PT}$ Symmetry} \ec

\vs{1cm}

\bc
{\it \"Ozlem Ye\c{s}ilta\c{s}$^{*}${\footnote {e-mail : yesiltas@gazi.edu.tr}   \\
$^{*}$Department of Physics, Faculty of Science, Gazi University,
06500 Ankara, Turkey\\
\vspace{.16cm}

}} \ec \vs{1cm}
\begin{abstract}
Two dimensional effective Hamiltonian for a massless Dirac electron interacting with a hyperbolic magnetic field is discussed within $\mathcal{PT}$ symmetry. Factorization method and polynomial procedures are used to solve Dirac equation for the constant Fermi velocity and the effective potential which is complex Scarf II potential. The more general effective Scarf II potential models are also obtained within pseudo-supersymmetry. Finally, an extension of Panella and Roy's work \cite{11} to the both $\mathcal{PT}$ symmetric and real Scarf II partner potentials are given using the position dependent Fermi velocity.
\end{abstract}
\noindent {\bf keyword:}  pt symmetry, Dirac electron \\


\section{Introduction}
The discovery of graphene \cite{Nov} and the massless Dirac character of the low
energy electrons due to the honeycomb lattice structure  have attracted much interest in physics. The intense study in the various effects in two-dimensional quantum systems was revived after the successful production of single-layer graphite (graphene) \cite{2, 3}. Recently, particle creation has become an observable effect in graphene physics  \cite{4, 5}. The low-energy electronic excitations in the graphene sheet in the presence of an external electromagnetic field can be described by the Dirac model \cite{6}. There have been many studies on various theoretical as well as experimental aspects of graphene. For example, a series of studies concerning the interaction of graphene electrons in perpendicular magnetic
fields \cite{7, 8, 9, Al}, position dependent Fermi velocity and analytical solutions \cite{10, 11, 12}, chiral symmetry breaking in graphene \cite{13}, exact solutions of the $(2+1)$ dimensional spacetime Dirac equation within minimal length \cite{14}. Moreover, methods of the super-symmetric quantum mechanics are used to obtain analytical solutions for the massless Dirac electrons in spherical molecules \cite{S2}, the Dirac equation in two dimensional curved space-time with Lie algebraic approach can be found in \cite{15}.

Symmetry in physics is one of the most vital topic attracted much interest in  every area of physics. $\mathcal{PT}$ symmetry  has acquired more interest which is always growing in recent years \cite{bender1, bender2, bender3, Bender, bender4}. $\mathcal{PT}$ symmetry is generated by the product of the parity and time reversal operators which are $\mathcal{P}$,  $ \mathcal{T}$ correspondingly. The parity operator is linear and has the effect of changing the sign of the position and momentum
operators $\mathcal{P}x\mathcal{P}=-x$, $\mathcal{P} p \mathcal{P}=-p$. The time reversal operator is anti-linear and has the effect $\mathcal{T}x\mathcal{T}=x$, $\mathcal{T} p \mathcal{T}=-p$, $\mathcal{T}i\mathcal{T}=-i$. The operator $\mathcal{T}$ is anti-linear because it changes the sign of $i$. If the $\mathcal{PT}$  symmetry of the Hamiltonian is unbroken; the eigenfunction of the operator $\mathcal{PT}$ is simultaneously an eigenstate of Hamiltonian $H$, i.e.  $[H,\mathcal{PT}]=0$. The five-dimensional approach to extend $\mathcal{T}$ and $\mathcal{PT}$ symmetry from non-relativistic to relativistic quantum mechanics was examined in  \cite{bender5}. On the other hand, in the relativistic area, exact solutions of the scalar Dirac equation transformed into a Klein-Gordon-like equations have attracted interest such as the Dirac equation for $\mathcal{PT}$ symmetric Hulthen potential is studied in \cite{16}, Darboux transformation for the one dimensional Dirac equation can be found in \cite{Roy}, the Dirac equation with $\mathcal{PT}$/non-$\mathcal{PT}$-symmetric potentials in the
presence of position-dependent mass is one of the interesting topics \cite{17, JD,Jia, 18,o}, the author has worked on a large class of non-Hermitian non-$\mathcal{PT}$-symmetric two-component Dirac Hamiltonians \cite{19}.

In this study, we have shown that the massless fermions can be modeled by Dirac-Weyl equation
within a general complex vector potential. Section 2 includes a brief explanation of the pseudo-supersymmetry. Section three is devoted to Dirac-Weyl equation and its $\mathcal{PT}$ symmetric properties and solvable effective super-partner potentials. Conclusions are involved in Section 4.
\section{Pseudo-Super-symmetry}
A subclass of non-Hermitian operators known as pseudo-Hermitian
operators \cite{m}. Here, the reality of the spectrum is
guaranteed if and only if there is a positive definite inner
product $\langle.,.\rangle_{+}$ on Hilbert space. This inner
product is expressed in terms of positive definite metric operator
$\eta$ which is invertible. Now, let us consider the intertwining
operator $\eta$ which is given by
\begin{equation}\label{15}
    \eta \mathcal{H}=H^{\dag}_{p} \eta
\end{equation}
where $\mathcal{H}=AB$, $A$ and $B$ are first order differential
operators, $H_{p}=BA$ is the partner Hamiltonian of $\mathcal{H}$
and  $H^{\dag}_{p}=BA$ is the adjoint of $\mathcal{H}$.
$\mathcal{H}$ is diagonalizable with a discrete spectrum such that
\begin{equation}\label{16}
    \mathcal{H}|\psi_{n},a\rangle=E_{n}|\psi_{n},a\rangle,~~~~ \mathcal{H^{~\dag}}|\phi_{n},a\rangle=E^{*}_{n}|\phi_{n},a\rangle
\end{equation}
\begin{equation}\label{17}
    \langle\phi_{m},b|\psi_{n},a\rangle=\delta_{mn}\delta_{ab},~~~~\sum_{n}\sum^{d_{n}}_{a=1}|\phi_{n},a\rangle\langle\psi_{n},a|=
    \sum_{n}\sum^{d_{n}}_{a=1}|\psi_{n},a\rangle\langle\phi_{n},a|=1
\end{equation}
where $a,b$ are degeneracy labels \cite{m}. $\mathcal{H}$ and $H_{p}$ are
related to each other with an intertwining operator $\eta_{1}$
\begin{equation}\label{18}
    \eta_{1}\mathcal{H}=H_{p}\eta_{1}
\end{equation}
and $H_{p}$ is $\eta_{2}$-pseudo-Hermitian if the relation below
is satisfied:
\begin{equation}\label{19}
    \eta_{2}H_{p}=H^{\dag}_{p}\eta_{2}.
\end{equation}
Then, there is an intertwining operator $\eta$ which is
$\eta=\eta_{2}\eta_{1}$  and $\eta \mathcal{H}=H^{\dag}_{p} \eta$.
\section{Dirac-Weyl Equation }
The low energy excitations about the K point in graphene are
described by the right-handed Dirac-Weyl equation which is
\begin{equation}\label{1}
    H_D=\upsilon_{F}(\vec{\sigma}\cdot \vec{p})
\end{equation}
where Fermi velocity $\upsilon_{F}=\frac{c}{300}$, $c$ is the
velocity of the light in the vacuum, $\vec{\sigma}=(\sigma_{x},
\sigma_{y})$ are the Pauli matrices and $\vec{p}=-i \hbar
(\frac{\partial}{\partial x}, \frac{\partial}{\partial y})$ is the
momentum operator given with respect to the center of the valley
at the corner of the Brillouin zone with wave vector $\vec{k}$. On
the other hand, the chirality in this case can be defined to be
the projection of the momentum on the direction of
pseudo-spin(internal degree of freedom plays a role of a
pseudospin) which is $\frac{\vec{\sigma} . \vec{p}}{|p|}$. If the
eigenvectors of the Hamiltonian in (\ref{1}) are $\Phi(x,y,t)$,
then, Dirac-Weyl equation turns into a time independent form as
\begin{equation}\label{2}
    \upsilon_{F} (\vec{\sigma}\cdot \vec{p})\Psi(x,y)=E\Psi(x,y)
\end{equation}
where $\Phi(x,y,t)=\Psi(x,y) e^{-\frac{iEt}{\hbar}}$. Plugging $\vec{p}\rightarrow \vec{p}+i \frac{e}{c} \vec{A}$
in (\ref{2}), we have
\begin{equation}\label{02}
    \upsilon_{F} (\vec{\sigma}\cdot (\vec{p}+i \frac{e}{c} \vec{A}))\Psi(x,y)=E\Psi(x,y).
\end{equation}
The Dirac electron-magnetic field interaction which is vertical to the plane of the graphene
can be written using the vector potential and the magnetic fields which are given by
\begin{equation}\label{3}
    \vec{A}(x)=(A_{x}, A_{y}, A_{z}), ~~~~\vec{B}=\vec{\nabla} \times \vec{A}.
\end{equation}
The eigenvector $\Psi(x,y)$ is taken as $\Psi(x,y)=(\phi_{1}(x,y),
\phi_{2}(x,y))^{T}$ where $T$ denotes the matrix transposition.
Let us choose the vector potential $\vec{A}$ which is
\begin{equation}\label{4}
    \vec{A}(x)=(0, A_{y}(x), 0),~~~~\vec{B}=(0, 0, B(x))
\end{equation}
where $B(x)=\frac{dA_{y}}{dx}$.
Now, the general Dirac operator  becomes
\begin{equation}\label{dt0}
H_D=\hbar\upsilon_F \left(
  \begin{array}{cc}
    0 & -i \partial_x-\partial_y+\frac{e}{c\hbar}A_{y}(x) \\
    -i \partial_x+\partial_y-\frac{e}{c\hbar}A_{y}(x) & 0 \\
  \end{array}
\right).
\end{equation}
We can give the effects of the $\mathcal{P}$ and $\mathcal{T}$ operators as below:
\begin{eqnarray}
  \mathcal{P}\psi(\vec{x},t) &=& \gamma^{0}\psi(-\vec{x},t) \\
 \mathcal{ T} \psi(\vec{x},t)&=& \psi^{*}(\vec{x},t)=\mathbf{K} \psi(\vec{x},t)
\end{eqnarray}
where $\mathbf{K}$ is the complex conjugation operator with an effect $\mathbf{K}: i\rightarrow -i$, the asterisk $^{*}$  superscript signifies the complex conjugate, $\gamma^{0}$ is one of the Dirac matrices $\gamma^{0}=\sigma^{3}$.
Let us discuss the $\mathcal{PT}$ transformed Dirac Hamiltonian matrix given by
\begin{equation}\label{dt}
    \mathcal{PT} H_D \mathcal{PT}=\hbar\upsilon_F \left(
                                    \begin{array}{cc}
                                      0 & i \partial_x+\partial_y-\frac{e}{c\hbar}A^{*}_{y}(-x) \\
                                      i \partial_x-\partial_y+\frac{e}{c\hbar}A^{*}_{y}(-x) & 0 \\
                                    \end{array}
                                  \right)
\end{equation}
where $\mathcal{P}^{2}=\mathcal{T}^{2}=1$ and $\mathcal{PT}=(\mathcal{PT})^{-1}$.
Hence, we note that  (\ref{dt0}) is not $\mathcal{PT}$ symmetric. Then, we can give
couple of differential equations given below:
\begin{equation}\label{6}
    \left(\frac{d}{dx}+k+i\frac{e}{c \hbar}
    A_{y}(x)\right)\psi_{2}(x)=\epsilon \psi_{1}(x),
\end{equation}
\begin{equation}\label{7}
  \left(-\frac{d}{dx}+k+i\frac{e}{c \hbar}
    A_{y}(x)\right)\psi_{1}(x)=\epsilon \psi_{2}(x)
\end{equation}
where $\epsilon=\frac{E}{\upsilon_{F}\hbar}$ and $\phi_{1}(x)=e^{i
k y} \psi_{1}(x)$, $\phi_{2}(x)=i e^{i k y} \psi_{2}(x)$. Using
(\ref{6}) and (\ref{7}) we obtain second- order  differential
equations
\begin{equation}\label{8}
  H_{1}\psi_{1}(x)= \left(-\frac{d^{2}}{dx^{2}}+\left(k+i\frac{e}{c \hbar}A_{y}\right)^{2}+i\frac{e}{c
    \hbar}A'_{y}(x)\right)\psi_{1}(x)=\epsilon^{2}\psi_{1}(x)
\end{equation}
\begin{equation}\label{9}
   H_{2}\psi_{2}(x)= \left(-\frac{d^{2}}{dx^{2}}+\left(k+i\frac{e}{c \hbar}A_{y}\right)^{2}-i\frac{e}{c
    \hbar}A'_{y}(x)\right)\psi_{2}(x)=\epsilon^{2}\psi_{2}(x).
\end{equation}
Thus, we have obtained partner Hamiltonians related to the
Dirac-Weyl equation and we may use
\begin{equation}\label{10}
    H_{1}=-\frac{d^{2}}{dx^{2}}+V_{1}(x)
\end{equation}
\begin{equation}\label{11}
 H_p=H_{2}=-\frac{d^{2}}{dx^{2}}+V_{2}(x).
\end{equation}
We may see from $H_1$ and $H_2$ that the super-potential $W(x)$  is in the form of
\begin{equation}\label{12}
    W(x)= k+i\frac{e}{c \hbar} A_{y}(x).
\end{equation}
Thus, one writes complex partner potentials $V_{1}(x)$ and
$V_{2}(x)$  using (\ref{12}) as
\begin{equation}\label{13}
    V_{1}(x)=W(x)^{2}+W'(x)=k^{2}-\frac{e^{2}A_{y}^{2}(x)}{c^{2}\hbar^{2}}+i\frac{e}{c\hbar}\left(2kA_{y}(x)+A'_{y}(x)\right)
\end{equation}
\begin{equation}\label{14}
    V_{2}(x)=W(x)^{2}-W'(x)=k^{2}-\frac{e^{2}A_{y}^{2}(x)}{c^{2}\hbar^{2}}+i\frac{e}{c\hbar}\left(2kA_{y}(x)-A'_{y}(x)\right).
\end{equation}
Also, we may look at the $\mathcal{C}$ symmetry of the one dimensional Dirac operator using (\ref{6}) and (\ref{7}). In this manner, the Dirac operator reads
\begin{equation}
\tilde{H}_D=\hbar \upsilon_{F}\left(
  \begin{array}{cc}
    0 & \partial_x+W(x) \\
    -\partial_x+W(x) & 0 \\
  \end{array}
\right).
\end{equation}
The operator $\mathcal{C}$ can be expressed by $\mathcal{C}=\gamma^{5}\mathrm{R}$ where $\gamma^{5}=i \gamma^{0}\gamma^{1}$, $\gamma^{1}=i \sigma^{1}$, $\mathrm{R}$ corresponds to the space reflection operator. One can show that $\mathcal{C} \tilde{H}_D \mathcal{C}^{-1}$ defines an anti- symmetry  $\mathcal{C} \tilde{H}_D \mathcal{C}^{-1}= - \tilde{H}_{D}(-e)$ if $W(-x)=W(x)$ or in other words $A_y(x)=-A_y(x)$.  Thus,
\begin{equation}
 \mathcal{C} \tilde{H}_D \mathcal{C}^{-1}=\hbar \upsilon_{F}\left(
  \begin{array}{cc}
    0 & -\partial_x-W(-x) \\
    \partial_x-W(-x) & 0 \\
  \end{array}
\right).
\end{equation}
The system under $\mathcal{CPT}$ symmetry behaves as $\mathcal{CPT} \tilde{H}_D \mathcal{CPT}^{-1}= - \tilde{H}_{D}(-e)$  if $A_y(x)$ is an even function.

\subsection{Constant Fermi velocity}
 Let us choose an even function which is $A_{y}(x)=A_{1}sech\mu x+A_{2}$, where $A_{1}$ and $A_{2}$ are real constants, then $\mathcal{PT}$ symmetric partner potentials
 $V_{1}(x)$ and $V_{2}(x)$ turn into
\begin{equation}\label{20}
    V_{1}(x)=-\frac{A^{2}_{1}e^{2}}{c^{2}\hbar^{2}}sech^{2}\mu x-i\frac{A_{1}e\mu}{c
    \hbar}sech\mu x\tanh \mu x
\end{equation}
and
\begin{equation}\label{21}
    V_{2}(x)=-\frac{A^{2}_{1}e^{2}}{c^{2}\hbar^{2}}sech^{2}\mu x+i\frac{A_{1}e\mu}{c
    \hbar}sech\mu x\tanh \mu x,
\end{equation}
where $A_{2}$ is taken as $A_{2}=\frac{c\hbar}{e}k$. It is seen that $V_1=V^{\dag}_2$. And the
magnetic field for this case is
\begin{equation}
    \vec{B}(x)=(0,0,- A_{1}\mu \sec h \mu x ~\tan h \mu x).
\end{equation}
Using (\ref{19}), we can obtain the operator $\eta_{2}(x)$ as
\begin{equation}\label{22}
    \eta_{2}(x)= \frac{d}{dx}+i\frac{A_{1}e}{c\hbar}sech\mu
    x.
\end{equation}
In order to find $\mathcal{H}$ which is the partner of $H_{1}$ and $H_2$, we give an ansatze
about the operator
\begin{equation}\label{23}
    \eta_{1}=\frac{d}{dx}+G(x)
\end{equation}
and we give the $\mathcal{H}$ with the unknown function $U(x)$ as
\begin{equation}\label{24}
    \mathcal{H}=-\frac{d^{2}}{dx^{2}}+U(x).
\end{equation}
Using $\eta_{1} \mathcal{H}=H_{2}\eta_{1}$, we obtain the
relations given below
\begin{eqnarray} \label{gx}
  2G'(x) &=& -U(x)-V_{1} sech^{2}\mu x+iV_{2} sech \mu x ~tanh \mu x \\
  G''(x) +U'(x)&=& -(U(x)+iV_{2}sech \mu x~ tanh \mu x+V_{1}sech^{2}\mu x)G(x) \label{gxx}
\end{eqnarray}
where $V_{1}=\frac{A^{2}_{1}e^{2}}{c^{2}\hbar^{2}}$,
$V_{2}=\frac{A_{1}e\mu}{c\hbar}$. $U(x)$ can be written using (\ref{gx}). Taking $G(x)$ as $G(x)=B_{1}tanh \mu x+B_{2}sech \mu x$, $B_1$ and $B_2$ are constants, we get
\begin{eqnarray}
  2B_{1}B_{2}-iV_{2}-B_{2}\mu &=&0  \label{111}\\
  B^{2}_{1}-B^{2}_{2}-B_1 \mu  -V_{1}&=& 0. \label{222}
\end{eqnarray}
$B_1$ and $B_2$ may be complex, imaginary or real constants. If we take them as $B_1=A+i H$, $B_2=J+i S$ and use (\ref{111}) and (\ref{222}), then, we find $J=H=0$ and
\begin{equation}\label{dort}
    \{(B_1); ( S)\}= \{\left(0, \mu, \frac{c \hbar-2A_1 e}{2c \hbar}, \frac{c \hbar+2A_1 e}{2c \hbar}\right);
    \left( - \frac{A_1 e}{c\hbar},  \frac{A_1 e}{c\hbar}, - \frac{\mu}{2},  \frac{\mu}{2} \right)\}.
\end{equation}
Then, we can obtain $U(x)$ as
\begin{equation}\label{25}
    U(x)= i(2S\mu+V_{2})sech\mu x \tanh \mu x-(\frac{V^{2}_{2}}{\mu^{2}}+2B_1 \mu)sec h^{2}\mu x.
\end{equation}
Let us give the $U(x)$ according to the set $ \{(B_1); ( S)\}$:
\begin{eqnarray}\label{250}
  (0; -\frac{A_1 e}{c \hbar}) &=& -i\frac{A_1 e \mu}{c\hbar}\sec h \mu x \tanh \mu x-\frac{A^{2}_{1}e^{2}}{c^{2}\hbar^{2}}sech^{2} \mu x \\\label{251}
  (\mu; \frac{A_1 e}{c \hbar}) &=& 3i\frac{A_1 e \mu}{c\hbar}\sec h \mu x \tanh \mu x-(\frac{A^{2}_{1}e^{2}}{c^{2}\hbar^{2}}+2\mu^{2})sech^{2} \mu x\\ \label{252}
  (\frac{c\hbar-2A_1 e}{2c\hbar}; -\frac{\mu}{2}) &=& i(-\mu^{2}+\frac{A_1 e \mu}{c\hbar})\sec h \mu x \tanh \mu x-(\frac{A^{2}_{1}e^{2}}{c^{2}\hbar^{2}}+\mu-\frac{2A_1 e \mu}{c\hbar})sech^{2} \mu x \\ \label{253}
 (\frac{c\hbar+2A_1 e}{2c\hbar}; -\frac{\mu}{2}) &=& i(\mu^{2}+\frac{A_1 e \mu}{c\hbar})sech\mu x\tanh \mu x-(\frac{A^{2}_{1}e^{2}}{c^{2}\hbar^{2}}+\mu+\frac{2A_1 e \mu}{c\hbar})sech^{2} \mu x. \label{254}
\end{eqnarray}
It is noted that (\ref{251}) agrees with the model given in \cite{roy} and (\ref{250}) equals to (\ref{20}). Next,
 we may give the operator $\eta_{1}(x)$ in the form,
\begin{equation}\label{291}
    \eta_{1}(x)=\frac{d}{dx}+ B_1 \tan h \mu x+i S ~sec h \mu x,
\end{equation}
and $\eta(x)$ becomes
\begin{equation}\label{292}
    \eta(x)=\left(\frac{d}{dx}+i\frac{A_{1}e}{c\hbar}sech\mu
    x\right)\left(\frac{d}{dx}+B_1 \tan h \mu x+iS~ \sec h \mu x\right).
\end{equation}
The hyperbolic complex potential given by the models (\ref{20}) and (\ref{21}) is known as complex Scarf II potential \cite{BSC, roy, que}. To find solutions of the Hamiltonian systems, we may use the polynomial solutions known as Nikiforov-Uvarov method \cite{NU1, NU2, NU3, NU4}. To start with, we shall use a variable transformation $z=\sinh\mu x$ in $\mathcal{H} \chi = \mathfrak{E} \chi$ with a general form of $U(x)$ as
\begin{equation}\label{293}
    U(x)= \textbf{A}_{1} sech^{2} \mu x+\textbf{A}_{2}sech \mu x \tanh \mu x.
\end{equation}
And we here use $\mathfrak{E}$ instead of $\epsilon^{2}$ in the eigenvalue equation which turns into
\begin{equation}\label{ev}
    \chi^{''}+\frac{z}{1+z^{2}}\chi^{'}+\frac{\tilde{\sigma}}{(1+z^{2})^{2}}\chi=0
\end{equation}
where
\begin{equation}
  \tilde{\sigma}=  \left(\frac{\mathfrak{E}}{\mu^{2}}(1+z^{2})-\frac{\textbf{A}_{1}}{\mu^{2}}-
    \frac{\textbf{A}_{2}z}{\mu^{2}}\right).
\end{equation}
Taking $\chi=\phi(z) y(z)$ in (\ref{ev}), then we have \cite{NU1}
\begin{equation}\label{ev1}
    y^{''}(z)+\left(2\frac{\phi^{'}}{\phi}+\frac{z}{1+z^{2}}\right)y^{'}(z)+\left(\frac{\phi^{''}}{\phi}+
    \frac{\phi^{'}}{\phi}\frac{z}{1+z^{2}}+\frac{1}{(1+z^{2})^{2}}\left(\frac{\mathfrak{E}}{\mu^{2}}(1+z^{2})-\frac{\textbf{A}_{1}}{\mu^{2}}-
    \frac{\textbf{A}_{2}z}{\mu^{2}}\right)\right)y(z)=0.
\end{equation}
The coefficient of $y^{'}$ has a role to bring a simplicity to (\ref{ev1}), then it is taken as a polynomial of degree at most 1. In that case we write \cite{NU1}
\begin{equation}\label{ev2}
    \frac{\phi^{'}(z)}{\phi(z)}=\frac{\pi(z)}{1+z^{2}}
\end{equation}
where
\begin{equation}\label{ev3}
    \pi(z)=\frac{1}{2}(\tau(z)-z),~~~~~\frac{\phi^{''}}{\phi}=\left(\frac{\pi}{1+z^{2}}\right)^{'}+\left(\frac{\pi}{(1+z^{2})}\right)^{2}.
\end{equation}
The final form of equation (\ref{ev1}) turns into
\begin{equation}\label{ev4}
    y^{''}(z)+\frac{\tau(z)}{1+z^{2}}y^{'}(z)+\frac{\bar{\sigma}(z)}{1+z^{2}}y(z)=0
\end{equation}
where
\begin{equation}\label{ev5}
  \bar{\sigma}(z)=  \left(\left(\frac{\mathfrak{E}}{\mu^{2}}(1+z^{2})-\frac{\textbf{A}_{1}}{\mu^{2}}-
    \frac{\textbf{A}_{2}z}{\mu^{2}}\right)+\pi^{2}(z)+\pi(z)(\tilde{\tau}-\sigma^{'})+\pi^{'}\sigma(z)\right).
\end{equation}
Because the polynomial $\bar{\sigma}$ is divisible by $\sigma(z)=1+z^{2}$ and $\bar{\sigma}(z)=\lambda \sigma(z)$, $\lambda$ should be a constant. Finally
(\ref{ev4}) becomes
\begin{equation}\label{ev6}
    \sigma(z)y^{''}+\tau(z)y^{'}+\lambda y=0.
\end{equation}
(\ref{ev6}) is known as hypergeometric type equation. To obtain eigenvalues we will use $\pi(z)$ \cite{NU1}
\begin{equation}\label{ev7}
    \pi(z)= \frac{\sigma^{'}(z)-\tilde{\tau}}{2}\pm \sqrt{(\frac{\sigma^{'}(z)-\tilde{\tau}}{2})^{2}-\tilde{\sigma}+k\sigma}
\end{equation}
where
\begin{equation}\label{k}
    k=\lambda-\pi^{'}(z).
\end{equation}
The properties of hypergeometric equation were investigated in \cite{NU1}. It is shown that every solution of hypergeometric equation when $\lambda\neq 0$ is the derivative of a solution of (\ref{ev6}). Thus we may write
\begin{equation}\label{la}
    \lambda=\lambda_{n}=-n \tau^{'}-\frac{n(n-1)}{2}\sigma^{''}
\end{equation}
and the solutions are given by
\begin{equation}\label{so}
    y_n(z)=\frac{N}{\rho(z)}(\sigma^{n}(z)\rho(z))^{n},~~~~n= 0, 1,...
\end{equation}
where
\begin{equation}\label{rho}
    (\sigma\rho)^{'}=\tau \rho.
\end{equation}
Additionally, there is a condition on $\rho$ such that $\tau(z)$ has to vanish at points $z\in (a,b)$ and $\tau(z)$ has a negative derivative $\tau^{'}(z)<0$. Due to the fact that $\pi(z)$ is a polynomial, the term under the square root in (\ref{ev7}) is necessarily the square of a polynomial. This happens if the discriminant of the term under the square root in (\ref{ev7}) is zero which leads to a quadratic equation for $k$. After defining $k$, we put $k$  in (\ref{ev7}). Thus, we get
\begin{equation}
\pi=\left\{
  \begin{array}{ll}
    \frac{z}{2}-\frac{1}{2\sqrt{2}}(\sqrt{-\sqrt{(1-4\mathcal{A}_{1})^{2}+16\mathcal{A}^{2}_{2})}+1-4\mathcal{A}_{1}} z+\\ \sqrt{-\sqrt{(1-4\mathcal{A}_{1})^{2}+16\mathcal{A}^{2}_{2}}-(1-4\mathcal{A}_{1})}), & \hbox{$k_1=\frac{1}{8}(-1-4\mathcal{A}_1+8\mathfrak{\bar{E}}-\sqrt{(4\mathcal{A}_1-1)^{2}}+16\mathcal{A}^{2}_2)$;} \\
    \frac{z}{2}-\frac{1}{2\sqrt{2}}(\sqrt{\sqrt{(1-4\mathcal{A}_{1})^{2}+16\mathcal{A}^{2}_{2})}+1-4\mathcal{A}_{1}} z+\\ \sqrt{\sqrt{(1-4\mathcal{A}_{1})^{2}+16\mathcal{A}^{2}_{2}}-(1-4\mathcal{A}_{1})}), & \hbox{$k_2=\frac{1}{8}(-1-4\mathcal{A}_1+8\mathfrak{\bar{E}}+\sqrt{(4\mathcal{A}_1-1)^{2}}+16\mathcal{A}^{2}_2)$.}
  \end{array}
\right.
\end{equation}
Here $\textbf{A}_{1,2}/\mu^{2}=\mathcal{A}_{1,2}$ and $\mathfrak{\bar{E}}=\mathfrak{E}/\mu^{2}$. Finally we obtain energy eigenvalues for each $k$ value as
\begin{equation}\label{energy1}
    \mathfrak{E}^{k_1}_n=-\mu^{2}\left(n+\frac{1}{2}-\frac{\sqrt{1-4\mathcal{A}_1-\sqrt{(1-4\mathcal{A}_1)^{2}+
    16\mathcal{A}^{2}_2}}}{2\sqrt{2}}\right)^{2}
\end{equation}
and
\begin{equation}\label{energy2}
    \mathfrak{E}^{k_2}_n=-\mu^{2}\left(n+\frac{1}{2}-\frac{\sqrt{1-4\mathcal{A}_1+\sqrt{(1-4\mathcal{A}_1)^{2}+
    16\mathcal{A}^{2}_2}}}{2\sqrt{2}}\right)^{2}.
\end{equation}
In both cases $\tau'<0$ is satisfied. Matching (\ref{293}) with (\ref{25}), we give
\begin{equation}\label{match1}
    \textbf{A}_{1}=\mu^{2}\mathcal{A}_{1}=-(\frac{V^{2}_{2}}{\mu^{2}}+2B_1 \mu)
\end{equation}
and
\begin{equation}\label{match1}
    \textbf{A}_{2}=\mu^{2}\mathcal{A}_{2}=i (V_2+2S\mu).
\end{equation}
Hence the spectrum of $\mathcal{H}$  for $k_1$ is given by
\begin{equation}\label{33}
      E^{k_1}_{n}= \pm i \mu^{2} \hbar \upsilon_{F} \left(n+\frac{1}{2}-\frac{\sqrt{1-4\mathcal{A}_1-\sqrt{(1-4\mathcal{A}_1)^{2}+16\mathcal{A}^{2}_2}}}{2\sqrt{2}}\right)
\end{equation}
and for $k_2$, we obtain
\begin{equation}\label{34}
      E^{k_2}_{n}= \pm i \mu^{2} \hbar \upsilon_{F} \left(n+\frac{1}{2}-\frac{\sqrt{1-4\mathcal{A}_1+\sqrt{(1-4\mathcal{A}_1)^{2}+16\mathcal{A}^{2}_2}}}{2\sqrt{2}}\right).
\end{equation}
We note that if we take the Fermi velocity as the imaginary Fermi velocity
$\upsilon_{F}\rightarrow i \upsilon_{F}$, we obtain the real
spectrum in both cases. Corresponding unnormalized solutions can be obtained using (\ref{so}), (\ref{rho}) as
\begin{equation}\label{wf}
    \chi^{k_1}_{n}(x)= N_{1} (\cosh \mu x)^{(\frac{1}{2}+\alpha_{1})}
    e^{-\beta_{1} Arc\tan(\sinh \mu x)} P^{(\alpha_{1}; 2\beta_{1})}_{n}(\sinh \mu x).
    \end{equation}
where
\begin{equation}\label{alb}
    \alpha_{1}=-\frac{\sqrt{1-4\mathcal{A}_1-\sqrt{(1-4\mathcal{A}_1)^{2}+16\mathcal{A}^{2}_{2}}}}{2\sqrt{2}},~~~~~
    \beta_{1}=\frac{\sqrt{-1+4A_1-\sqrt{(1-4A_{1})^{2}+16A^{2}_{2}}}}{2\sqrt{2}},
\end{equation}
and
\begin{equation}\label{wf}
    \chi^{k_2}_{n}(x)= N_{2} (\cosh \mu x)^{(\frac{1}{2}+\alpha_{2})}
    e^{-\beta_{2} Arc\tan(\sinh \mu x)} P^{(\alpha_{2}; 2\beta_{2})}_{n}(\sinh \mu x).
    \end{equation}
where
\begin{equation}\label{alb}
    \alpha_{2}=-\frac{\sqrt{1-4\mathcal{A}_1+\sqrt{(1-4\mathcal{A}_1)^{2}+16\mathcal{A}^{2}_{2}}}}{2\sqrt{2}},~~~~~
    \beta_{2}=\frac{\sqrt{-1+4A_1+\sqrt{(1-4A_{1})^{2}+16A^{2}_{2}}}}{2\sqrt{2}}.
\end{equation}
We note that $P^{(a;b)}_n(y)$ stands for the Jacobi polynomials, $N_1, N_2$ are the normalization constants. There are some properties that can be made out from the intertwining relations. If we assume that the spectrum of $\mathcal{H}$ ($H_2$) is known, then its partner $H_2$ ($\mathcal{H}$) will have the same spectrum except the ground state. For instance, for (\ref{250}) and $V_2(x)$ we may give the superpotential function  as
\begin{equation}\label{supp}
    \textbf{W}(x)=-i\frac{A_1 e}{c\hbar}sech \mu x,
\end{equation}
now the potential functions are given below
\begin{equation}\label{hler}
    U(x)=\textbf{W}^{2}(x)-\textbf{W}'(x)
\end{equation}
and
\begin{equation}\label{hle}
    V_2(x)=\textbf{W}^{2}(x)+\textbf{W}'(x).
\end{equation}
If we remind that the Hamiltonians $\mathcal{H}$ and $H_2$ were linked by the intertwining (pseudo-supersymmetric) transformations, we may write
\begin{equation}\label{tra}
    \mathcal{H}\eta^{-1}=\eta^{-1}H_2.
\end{equation}
If we start with a spectrum of $\mathcal{H}$ which is known and supposing that the ground-state of $H_2$ vanishes
\begin{equation}
    \eta^{-1}_1 \chi_{0}=0.
\end{equation}
Then we have $E_{0}=0$. And the spectrum relationship between these two Hamiltonians is
\begin{equation}\label{sp}
    E_{n}=E_{2,n-1}
\end{equation}
and the eigenfunctions are given as
\begin{equation}
    \psi_{2, n-1}=\frac{1}{\sqrt{E_{n}}} \eta^{-1}_{1} \chi_{n},~~~~n=1,2,...
\end{equation}
\subsection{Position dependent Fermi velocity and deformed potential models}
In \cite{4}, electronic transport is studied in one-dimensional hetero-structures using Dirac equation. The author used a trivial replacement $ \upsilon_{F}\rightarrow  \upsilon_{F}(x)$ and gave a Hermitian operator
\begin{equation}\label{ho}
    h=\sqrt{ \upsilon_{F}(x)}\sigma_{x}\frac{\hbar}{i}\frac{d}{dx}\sqrt{ \upsilon_{F}(x)}.
\end{equation}
According to (\ref{ho}), let us give (\ref{1}) using the position dependent Fermi velocity,
\begin{equation}\label{Ho}
    \mathbb{H}_D=\sqrt{ \upsilon_{F}(x)}(\vec{\sigma}\cdot(\vec{p}\sqrt{ \upsilon_{F}(x)}+i\frac{e}{c}\vec{A})).
\end{equation}
Using $\mathbb{H_D} \Phi(\vec{x},t)=0$ and $\Phi(\vec{x})=\Psi(\vec{x}) e^{-i\frac{E t}{\hbar}}$, we can obtain two couple of non-Hermitian Dirac Hamiltonian operators:
\begin{equation}\label{H1}
   \mathbb{H}_1=- \upsilon^{2}(x)\frac{d^{2}}{dx^{2}}- \upsilon_{F}(x) \upsilon'(x)\frac{d}{dx}+ \upsilon(x) \upsilon'(x)\left(k+i\frac{e}{\hbar}A_{y}(x)\right)+
 \upsilon^{2}(x) \left(k+i\frac{e}{\hbar}A^{2}_{y}(x)\right)^{2}+ i\frac{e}{\hbar}\upsilon^{2}(x)A'_{y}(x)
\end{equation}
and
\begin{equation}\label{H2}
   \mathbb{H}_2=- \upsilon^{2}(x)\frac{d^{2}}{dx^{2}}- \upsilon(x) \upsilon'(x)\frac{d}{dx}- \upsilon(x) \upsilon'(x)\left(k+i\frac{e}{\hbar}A_{y}(x)\right)+
 \upsilon^{2}(x) \left(k+i\frac{e}{\hbar}A^{2}_{y}(x)\right)^{2}- i\frac{e}{\hbar}\upsilon^{2}(x)A'_{y}(x)
\end{equation}
where $\upsilon_{F}(x)=\upsilon_{F} \upsilon(x)$, $\upsilon_{F}$ is a constant and the eigenvalue equations can be given by $\mathbb{H}_{j}\Psi_j=\epsilon^{2}\Psi_j$, $\epsilon=\frac{E}{\upsilon_{F}}$, $j=1,2$ and $\Psi(\vec{x})=e^{iky}\sqrt{\upsilon_{F}(x)}[\psi_1(x)~~~i\psi_2(x)]^{T}$. We note that  we may choose $A_{y}(x)\rightarrow\sqrt{\upsilon_{F}(x)}A_{y}(x)$. The couple of Hamiltonians (\ref{H1}) and (\ref{H2}) can be introduced as,
\begin{eqnarray}\label{eff1}
  \mathbb{H}_1 &=& - \upsilon^{2}(x)\frac{d^{2}}{dx^{2}}- \upsilon(x) \upsilon'(x)\frac{d}{dx}+\mathbb{W}(x)^{2}+(\upsilon(x)\mathbb{W}(x))' \\\label{eff2}
  \mathbb{H}_2 &=& - \upsilon^{2}(x)\frac{d^{2}}{dx^{2}}- \upsilon(x) \upsilon'(x)\frac{d}{dx}+\mathbb{W}(x)^{2}-(\upsilon(x)\mathbb{W}(x))'
\end{eqnarray}
where
\begin{equation}\label{sw}
    \mathbb{W}(x)=\upsilon(x)(k+i\frac{e}{\hbar}A_{y}(x)).
\end{equation}
We also indicate that the Hamiltonians given by (\ref{H1}) and (\ref{H2}) are transformed into a form given by
\begin{equation}\label{hj}
\begin{split}
    h_j=- \upsilon^{2} (x)\frac{d^{2}}{dx^{2}}- 2\upsilon (x) \upsilon' (x)\frac{d}{dx}+s_{j} \upsilon (x) \upsilon' (x)\left(k+i\frac{e}{\hbar}A_{y}(x)\right)&+\\
 \upsilon^{2} (x) \left(k+i\frac{e}{\hbar}A^{2}_{y}(x)\right)^{2}+s_{j}~ i\frac{e}{\hbar}\upsilon^{2} (x)A'_{y}(x)-
 \frac{1}{2}\upsilon (x)\upsilon ''(x)-\frac{\upsilon '^{2}}{4}
\end{split}
\end{equation}
where $\psi_{j}(x)=e^{\int^{x}\frac{\upsilon^{'} (y)}{2\upsilon (y)}dy}\phi_j(x)$ and we use
\begin{equation}\label{sj}
    s_{j}=\left\{
               \begin{array}{ll}
                 +, & \hbox{$j=1$;} \\
                 -, & \hbox{$j=2$.}
               \end{array}
             \right.
\end{equation}
The transformed Hamiltonian in (\ref{hj}) can also be expressed as
\begin{equation}\label{hj1}
    h_j=- \upsilon^{2} (x)\frac{d^{2}}{dx^{2}}- 2\upsilon (x) \upsilon' (x)\frac{d}{dx}+\tilde{V}_{eff,j}(x)
\end{equation}
where
\begin{equation}\label{vf}
    \tilde{V}_{eff,j}(x)=V_{eff,j}(x)+\rho(\upsilon)
\end{equation}
\begin{eqnarray}\label{eff1}
  V_{eff, 1}(x) &=& \mathbb{W}^{2}(x)+ (\upsilon (x)\mathbb{W}(x))'\\\label{eff2}
  V_{eff, 2}(x) &=& \mathbb{W}^{2}(x)- (\upsilon (x)\mathbb{W}(x))'
\end{eqnarray}
and
\begin{equation}\label{up}
    \rho(\upsilon)=-\frac{\upsilon(x)\upsilon''(x)}{2}-\frac{\upsilon'(x)^{2}}{4}.
\end{equation}
Here, $\rho(\upsilon)$ is known as the pseudo-potential term \cite{nieto}. In \cite{tha}, the position dependent mass Schr\"{o}dinger equation was reinterpreted as a deformed Schr\"{o}dinger equation using a momentum operator that reads \\ $\pi=\sqrt{f(\alpha,x)}p\sqrt{f(\alpha,x)}$  instead of the momentum operator $p=-i\frac{d}{dx}$. The potential in this work has a form $V(x)=W(x)^2-f(\alpha,x)W(x)'$, $W(x)$ is the superpotential, $\alpha$ is a constant. But in our work, (\ref{eff1}) and (\ref{eff2}) seem to be  different models as another one given in \cite{dutra}. Let us now give an ansatze for both $\upsilon(x)$ and $A_y(x)$ in (\ref{hj1}) to obtain a solvable Scarf II potential models that may be real or complex.
\subsubsection{real effective potentials}
We seek for a hyperbolic Scarf II potential model for (\ref{vf}). Then, we may give an ansatze for each unknown function in (\ref{vf})  which are
\begin{eqnarray}\label{up}
  \upsilon(x) &=&  \beta+\alpha \sinh \mu x \\ \label{A}
  A_y(x) &=& A_0\tanh \mu x+i A_1 sech \mu x+A_2+i A_3 \\
  A_3 &=& \frac{i A_{2} e+\hbar k}{e} \label{A3}
\end{eqnarray}
where $A_0, A_1, A_2, A_3$ are constants. Using  (\ref{up}) and (\ref{A})  in  (\ref{vf}), one obtains
\begin{equation}\label{v2e}
\begin{split}
   \tilde{V}_{eff,2}(x)= \frac{A_1 e \alpha \beta \mu}{\hbar}-\frac{\alpha^{2}\mu^{2}}{4}+(\frac{A^{2}_{1}e^{2}\beta^{2}}{\hbar^{2}}-i\frac{A_0 e \beta^{2}\mu}{\hbar})sech^{2}\mu x+(\frac{A_1 e \alpha^{2}\mu}{\hbar}&-\\ i\frac{A_0 e \alpha \beta \mu}{\hbar}-\frac{\alpha \beta \mu^{2}}{2})\sinh \mu
   x-(i\frac{A_0e\alpha^{2}\mu}{\hbar}+\frac{3\alpha^{2}\mu^{2}}{4})\sinh^{2}\mu x+2i(\frac{A^{2}_{1}e^{2}\alpha\beta}{i\hbar^{2}}-\frac{A_{1}A_{0}e^{2}\beta^{2}}{\hbar^{2}}-\\ \frac{A_0 e\alpha\beta\mu}{\hbar}-\frac{A_1e\beta^{2}\mu}{2i\hbar})sech\mu x\tanh \mu x-\frac{e^{2}}{\hbar^{2}}(((A_1\alpha-iA_0 \beta)^{2}-
   2iA_0 A_1 \alpha\beta)-i\frac{A_0 e\alpha^{2}\mu}{\hbar}-\frac{2A_1e\alpha\beta\mu}{\hbar})\tanh^{2}\mu x
   +\\ 2i(-\frac{A_0 A_1 e^{2}\alpha^{2}}{\hbar^{2}}-\frac{A^{2}_{0}e^{2}\alpha\beta}{i\hbar^{2}}-
   \frac{A_1e\alpha^{2}\mu}{2\hbar})\sinh\mu x\tanh^{2} \mu x-\frac{A^{2}_{0}e^{2}\alpha^{2}}{\hbar^{2}}\sinh^{2}\mu x\tanh^{2}\mu x.
\end{split}
\end{equation}
In order to get a solvable effective potential model,  we will simplify (\ref{v2e}). Then, one can obtain
\begin{equation}
\begin{split}\label{v2e0}
 \tilde{V}_{eff, 1}(x) = -(\beta^{2}+\alpha^{2})\mu+(\frac{3\alpha^{2}\mu^{2}}{4}+\beta^{2}\mu^{2}+\frac{\beta^{4}\mu^{2}}{4\alpha^{2}})sech^{2}\mu x
  -\alpha^{2}\mu^{2}\sinh^{2}\mu x- \\ \frac{\alpha\beta}{2}\sinh\mu x(1+\tanh^{2}\mu x)-\beta(\alpha+\frac{\beta^{2}}{2\alpha})\mu^{2}sech\mu x\tanh\mu x
\end{split}
\end{equation}

and
\begin{equation}
\begin{split}\label{v2e00}
 \tilde{V}_{eff, 2}(x) =\left(  -\frac{\alpha^{2}}{4}+\frac{\beta^{4}}{4\alpha^{2}}\right)\mu^{2}sech^{2}\mu x
  +\frac{\beta \mu^{2} }{2}\left(\alpha+\frac{\beta^{2}}{\alpha}\right)sech\mu x\tanh\mu x.
\end{split}
\end{equation}
Here, the constants $A_0, A_1, A_2$ must satisfy some constraints which are
\begin{eqnarray}\label{v2e01}
  \{(A_0; A_1)\} &=& \{\frac{i\hbar\mu}{2e}; -\frac{\hbar\beta\mu}{2e\alpha}\} \\ \label{v2e2}
  \{(A_0; A_1)\} &=& \{-\frac{3i\hbar\mu}{2e}; \frac{5\hbar\beta\mu}{6e\alpha}\}.
\end{eqnarray}
We note that one of the constraints (\ref{v2e01}) is used to get (\ref{v2e0}) and (\ref{v2e00}).
\subsubsection{complex effective potentials}
In this case we will use complex $\upsilon(x)$ in $(\ref{hj})$ and take $A_y(x) \rightarrow i A_y(x)$ which makes the Hamiltonian $h_j$ real and Hermitian. This may also mean that we use $\vec{p}\rightarrow \vec{p}- \frac{e}{c} \vec{A}$ in (\ref{02}). Now we give the ansazte for both $\upsilon(x)$ and $A_y(x)$ as below,
\begin{eqnarray}\label{son}
  \upsilon(x) &=&  i\beta+\alpha \sinh \mu x \\ \label{sonn1}
  A_y(x) &=& A_0\tanh \mu x+i A_1 sech \mu x+A_2+i A_3 \\
  A_3 &=& \frac{i (A_{2} e-\hbar k)}{e}.
\end{eqnarray}
We  remind that we will use (\ref{son}) and (\ref{sonn1}) in $\tilde{V}_{eff,1}(x)$ and $\tilde{V}_{eff,2}(x)$. Hence we obtain
\begin{equation}\label{v2ee3}
\begin{split}
   \tilde{V}_{eff,2}(x)= -\frac{A_1 e \alpha \beta \mu}{\hbar}-\frac{\alpha^{2}\mu^{2}}{4}+(\frac{A^{2}_{1}e^{2}\beta^{2}}{\hbar^{2}}-\frac{A_0 e \beta^{2}\mu}{\hbar})sech^{2}\mu x+(i\frac{A_1 e \alpha^{2}\mu}{\hbar}&+\\ \frac{A_0 e \alpha \beta \mu}{\hbar}-\frac{\alpha \beta \mu^{2}}{2})\sinh \mu
   x+(\frac{A_0e\alpha^{2}\mu}{\hbar}-\frac{3\alpha^{2}\mu^{2}}{4})\sinh^{2}\mu x+2i(-\frac{A^{2}_{1}e^{2}\alpha\beta}{\hbar^{2}}-\frac{A_{1}A_{0}e^{2}\beta^{2}}{\hbar^{2}}+\\ \frac{A_0 e\alpha\beta\mu}{\hbar}+\frac{A_1e\beta^{2}\mu}{2\hbar})sech\mu x\tanh \mu x-\frac{e^{2}}{\hbar^{2}}((-(A_1\alpha+A_0 \beta)^{2}-
   2A_0 A_1 \alpha\beta)+\frac{A_0 e\alpha^{2}\mu}{\hbar}+\frac{2A_1e\alpha\beta\mu}{\hbar})\tanh^{2}\mu x
   +\\ 2i(\frac{A_0 A_1 e^{2}\alpha^{2}}{\hbar^{2}}+\frac{A^{2}_{0}e^{2}\alpha\beta}{\hbar^{2}}-
   \frac{A_1e\alpha^{2}\mu}{2\hbar})\sinh\mu x\tanh^{2} \mu x+\frac{A^{2}_{0}e^{2}\alpha^{2}}{\hbar^{2}}\sinh^{2}\mu x\tanh^{2}\mu x.
\end{split}
\end{equation}
Simplifying (\ref{v2ee3}) leads to
\begin{equation}\label{son1}
\begin{split}
    \tilde{V}_{eff,1}(x)=(\beta^{2}-\alpha^{2})\mu+(\frac{3\alpha^{2}\mu^{2}}{4}-\beta^{2}\mu^{2}+\frac{\beta^{4}\mu^{2}}{4\alpha^{2}})sech^{2}\mu x
  -\alpha^{2}\mu^{2}\sinh^{2}\mu x- \\i \frac{\alpha\beta}{2}\mu^{2}\sinh\mu x(1+\tanh^{2}\mu x)+i\beta(-\alpha+\frac{\beta^{2}}{2\alpha})\mu^{2}sech\mu x\tanh\mu x
  \end{split}
\end{equation}
and
\begin{equation}\label{son2}
    \tilde{V}_{eff,2}(x)=\frac{\mu^{2}}{4}\left(\frac{\beta^{4}}{\alpha^{2}}-\alpha^{2}\right)sech^{2}\mu x+i\frac{\beta \mu^{2}}{2}\left(\alpha-\frac{\beta^{2}}{\alpha}\right)sech\mu x\tanh\mu x,
\end{equation}
where $A_0$ and $A_1$ should satisfy
\begin{eqnarray}\label{v2e5}
  \{(A_0; A_1)\} &=& \{\frac{\hbar\mu}{2e}; -\frac{\hbar\beta\mu}{2e\alpha}\} \\ \label{v2e2}
  \{(A_0; A_1)\} &=& \{-\frac{3\hbar\mu}{2e}; \frac{5\hbar\beta\mu}{6e\alpha}\},
\end{eqnarray}
and (\ref{v2e5}) is used to obtain (\ref{son1}) and (\ref{son2}).

\newpage

\section{Conclusion}
In the current paper, we have considered the Dirac-Weyl equation for the massless fermions in $\mathcal{PT}$ symmetric quantum mechanics. The aim of this work  is to show the connection of the massless Dirac–-Weyl equation and $\mathcal{PT}$ symmetric quantum mechanics. Hence, we have discussed the parity $\mathcal{P}$, time $\mathcal{T}$, $\mathcal{C}$, $\mathcal{PT}$ and $\mathcal{CPT}$ symmetries of the Dirac matrix operator. The Dirac Hamiltonian system including a pair of first order differential equations are  given in a decoupled second order differential equations and effective Dirac Hamiltonians are expressed in terms of the complex superpotential. Then, we have obtained  solvable potential models, complex Scarf II partner potentials, using the ansatze for the unknown  function that is vector potential function. The properties of the intertwining relations in pseudo-supersymmetric quantum mechanics are then used  to obtain a third Hamiltonian linked with those partner Hamiltonians  where we have also formed the intertwining operators $\eta_1, \eta_2$. We have seen that $\mathcal{H}$ given by (\ref{24}) may be the one of the Hamiltonian whose possible effective potentials are given by (\ref{250}), (\ref{251}), (\ref{252}) and (\ref{253}). Hereafter we have employed the polynomial procedures called as Nikiforov-Uvarov method to find the solutions of $\mathcal{H}$ and obtained the energy spectra and eigenfunctions. It is seen that the energy spectrum is imaginary in case of a real Fermi velocity while the energies of the Hamiltonian $\mathcal{H}$ are real if $\upsilon_{F}$ is taken as $\upsilon_F \rightarrow i\upsilon_F$. It is also pointed out that $H_2$ and $\mathcal{H}$ share the same spectrum except the ground-state level. Afterwards we have taken the Fermi velocity as a position dependent function according to the results of \cite{10} and it is observed that real position dependent Fermi velocity function leads to a real effective potential even if the vector potential is taken as a complex function. But complex Fermi velocity function leads to complex effective potential model. We have also seen that the partner effective potential of the $\tilde{V}_{2,eff}(x)$ that is called as 
$\tilde{V}_{1,eff}(x)$ is obtained as non-solvable potential model in both cases. Consequently it is seen that more general models may be generated through the position dependent Fermi velocity function. Finally, future studies may be devoted to the symmetries and their classification of the effective models of the Dirac systems (\ref{Ho}).

\newpage


\begin{thebibliography}{99}
\bibitem{Nov} Novoselov K S, Geim A K, Morozov S M, Zhang Y,
Dubonos S V, Grigorieva I V and Firsov A A 2004 Science
306 666.

\bibitem{2} Castro Neto A H, Guinea F, Peres N M, Novoselov K S and
Geim A K 2009 Rev. Mod. Phys. 81 109.
\bibitem{3}Gusynin V P and Sharapov S G 2008 Phys. Rev. Lett.
95 146801.
\bibitem{4} N. M. R. Peres, Rev. Mod. Phys. 82, 2673
(2010).
\bibitem{5} D. Das Sarma, S. Adam, E. H. Hwang, and E. Rossi, Rev.
Mod. Phys. 83, 407 (2011).
\bibitem{6} D. Allor, T. D. Cohen, and D. A. McGady, Phys. Rev. D
78, 096009 (2008).

\bibitem{7} S. Kuru, J. Negro, L. M.  Nieto, J. Phys.-Cond. Matt., 21(45) 455305 2009.
\bibitem{8} J. Zhu, S. M. Badalyan, F. Peeters, Phys. Rev. Lett. 109, 256602 (2012).
\bibitem{9} S. Liu, A. Nurbawono, N. Guo and C. Zhang, 2013 J. Phys.: Condens. Matter 25 395302.
\bibitem{Al} A. D. Alhaidari et al, Eur. Phys. J. B (2013) 86 73.

\bibitem{10} N.M.R. Peres, J. Phys.: Condens. Matter 21 (2009) 095501.
\bibitem{11} O. Panella and P. Roy, Physics Letters A 376 (2012) 2580–2583.
\bibitem{12} O. Mustafa, Cent. Eur. J. Phys. 11(4) 480 2013.

\bibitem{13} Y. Araki, Ann. Phys. 326 2011 1408.
\bibitem{14} L. Menculini, O. Panella and P. Roy, 87 065017 (2013).

\bibitem{S2} V. Jakubsky, S. Kuru, J.  Negro, S. Tristao, J. Phys.-Cond. Matt. 25(16) 165301 2013.

\bibitem{15} V. Jakubsky, Ann. Phys.  331 216 2013.


\bibitem{bender1} C. M. Bender and S. Boettcher, Phys. Rev. Lett. \textbf{80},  5243 (1998).
\bibitem{bender2} For a review, see C. M. Bender Rep. Prog. Phys. \textbf{70},  947 (2007).
\bibitem{bender3} C. M. Bender and H. F. Jones, Phys. Rev. A \textbf{85}, 052118 (2012).
\bibitem{Bender} C. M. Bender, Czec. J. Phys. \textbf{54}, 1027 (2004).
\bibitem{bender4} C. M. Bender, D. C. Brody and H. F. Jones  Phys. Rev. Lett. \textbf{89},
270401 2002; \textbf{92}, 119902 (2002).
\bibitem{bender5} Carl M. Bender and Philip D. Mannheim, Phys. Rev. D 84 129902 (2011).


\bibitem{16} E\~{g}rifes H, Sever R, Phys. Lett. A  344 117 2005.
\bibitem{Roy} A. Sinha and P. Roy, Int.J.Mod.Phys. A21 (2006) 5807-5822.

\bibitem{17} C.-S. Jia and A de S. Dutra, J. Phys. A: Math. Gen. 39 11877 2006.
\bibitem{JD} C.-S. Jia and A de S. Dutra, Ann. Phys. 323 566 2008.
\bibitem{Jia} Chun-Sheng Jia et al, Int. J. Theor. Phys. 47, 2513 (2008).

\bibitem{18} L. B. Castro, Phys.Lett. A375 (2011).
\bibitem{o} \"{O}. Ye\c{s}ilta\c{s}, J. Phys. A:Math. Theor. 46 015302 2013.
\bibitem{19} A.D. Alhaidari, Phys. Lett. A 377  2003 2013.
\bibitem{m} A. Mostafazadeh, J. Math. Phys.  43(1) 205 2002;  43(5)  2814 2002; 43(8) 3944 2002.
\bibitem{BSC} B. Bagchi, S. Mallik, C. Quesne, Int. J. Mod. Phys. A 16  2859 2001.
\bibitem{roy} R. Roychoudhury and P. Roy, Phys. Lett. A, 361 291 2007.
\bibitem{que} B. Bagchi and C. Quesne, Phys. Lett. A 300 26 2002.
\bibitem{NU1} A. F. Nikiforov, V. B. Uvarov, Special functions of mathematical physics:
a unified introduction with applications, Boston, MA: Birkhauser, 1988.
\bibitem{NU2}  F. B\"{u}y\"{u}kkili\c{c}, H. E\~{g}rifes and D. Demirhan, Theo. Chem. Acc. 98 192 1997.
\bibitem{NU3}  H. E\~{g}rifes, D. Demirhan and F. B\"{u}y\"{u}kkili\c{c}, Phys. Scr. 60(3) 195 1999.
\bibitem{NU4}  S. M. Ikhdair and R. Sever, App. Math. and Comp.  218(20)   10082  2012.


\bibitem{ahmed} Z. Ahmed, Phys. Lett. A 282 343 2001.
\bibitem{ahmed1} Z. Ahmed, Phys. Lett. A 287 295 2001, Addendum.


\bibitem{nieto} A. Ganguly and L M Nieto, J. Phys. A: Math. Theor. 40  7265 2007.
\bibitem{tha} B. Bagchi, A. Banerjee, C. Quesne and V. M. Tkachuk, J. Phys. A: Math. Gen. 38  2929 2005.
\bibitem{dutra} A. De S. Dutra,  M. Hott and C. A. S. Almeida , Europhys. Lett. 62 8 2003.

\end{thebibliography}
\end{document}